\documentclass[11pt,preprintnumbers,aps,amssymb,nofootinbib,amsmath,superscriptaddress]
{revtex4}
\usepackage{epsfig,epsf}
\usepackage{bm} 
\usepackage{color} 
\newcommand{\beq}{\begin{equation}}
\newcommand{\beql}[1]{\begin{equation}\label{#1}}
\newcommand{\eeq}{\end{equation}}
\def\bal#1\gal{\begin{align}#1\end{align}}
\newcommand{\ball}[1]{\bal\label{#1}}
\newcommand{\eq}[1]{(\ref{#1})}
\newcommand{\fig}[1]{Fig.~\ref{#1}}
\renewcommand{\sec}[1]{Sec.~\ref{#1}}
\newcounter{topiccounter}
\renewcommand{\b}[1]{{\bm #1}} 
\newcommand{\unit}[1]{\hat {{\bm #1}}} 

\newcommand{\as}{\alpha_s}

\newcommand{\e}{\varepsilon}

\begin{document}


\title{Classical electromagnetic fields from quantum sources in heavy-ion collisions}

\author{Robert Holliday}
\author{Ryan McCarty}
\author{Balthazar Peroutka}
\author{Kirill Tuchin}

\affiliation{Department of Physics and Astronomy, Iowa State University, Ames, Iowa, 50011, USA}

\date{\today}


\begin{abstract}

Electromagnetic fields are generated in high energy nuclear collisions by spectator valence protons. These fields are traditionally computed by integrating the Maxwell equations with point sources. One might expect that such an approach is valid at distances much larger than the proton size and thus such a classical approach should work well for almost the entire interaction region in the case of heavy nuclei. We argue that, in fact, the contrary is true: due to the quantum diffusion of the proton  wave function, the classical approximation breaks down at distances of the order of the system size.
We compute the electromagnetic field created by a charged particle described initially as a Gaussian wave packet of width 1~fm and evolving in vacuum according to the Klein-Gordon equation. We completely neglect the medium effects. We show that the dynamics, magnitude and even sign of the electromagnetic field created by classical and quantum sources are different.

\end{abstract}

\maketitle

\section{Introduction}\label{sec:intr}

Highly intense electromagnetic fields are created in relativistic  nuclear collisions. The first possible phenomenological manifestation of these fields in proton-proton collisions was proposed in 1988 by Ambjorn and Olesen \cite{Ambjorn:1988tm}, who suggested that they may cause condensation of $W$-bosons. Recently, it has been realized that the electromagnetic fields may get frozen into the Quark Gluon Plasma produced in thess relativistic heavy-ion collisions \cite{Tuchin:2013apa}, which has numerous  phenomenological consequences \cite{Tuchin:2013ie,Huang:2015oca}. How strong  the effect of the electromagnetic field is on observable quantities depends on the strength of the field at the time of collision and on its subsequent time-evolution. This information is encoded in Maxwell's equations, which can be solved for a given distribution of electromagnetic currents. 

All calculations of the electromagnetic fields thus far assumed that the sources are electrically charged point particles counter-propagating along the straight lines at a distance $b$ away from each other \cite{Skokov:2009qp,Tuchin:2010vs,Bzdak:2011yy,Voronyuk:2011jd,Deng:2012pc,Bloczynski:2012en,Zakharov:2014dia}. The resulting field strength is inversely proportional to the square of the impact parameter $b$. At small $b$ the classical approximation breaks down, which is manifested by the divergence of the field strength. One can readily fix this problem by imposing an ultra-violet cutoff, thereby effectively converting the point charges into hard spheres. \cite{Bzdak:2011yy,Bloczynski:2012en}. However, this prescription leads to a significant variation in field strength estimates. Since the largest contribution to the electromagnetic fields arises at short distances,  where the classical approximation breaks down, a consistent calculation requires the full quantum treatment of the electromagnetic currents. Such a treatment constitutes the main subject of the present article. The classical approximation is expected to break down at distances $b\sim 1/m$, because a particle's position cannot be localized better than its Compton wave length. In fact, we argue that this happens at much larger distances.

In the quantum approach, the colliding nucleons are described by a wave function satisfying the Dirac equation. Since the charge distribution in quantum treatment is not singular, the strength of the electromagnetic field naturally saturates in the UV limit.  As in the classical case, we employ the eikonal approximation \cite{Skokov:2009qp,Tuchin:2010vs,Bzdak:2011yy,Voronyuk:2011jd,Deng:2012pc,Bloczynski:2012en,Zakharov:2014dia}, assuming that nucleons move along the straight lines thus neglecting their deflection due to the external fields. This approximation breaks down at very large momentum transfer corresponding to very small $b\sim \alpha_s/m\gamma$, where $m$ is the nucleon mass and $\gamma$ is the collisions energy in units of $m$. The principal practical difference between the quantum and classical approaches is the quantum diffusion effect of the nucleon wave function. In contrast to the classical picture where the charge moves as a whole at nearly the speed of light, quantum diffusion generates a sideway flow of  current. As a consequence, the electromagnetic field stays longer in the interaction region. This effect does not require any medium and is present even in vacuum. Therefore, in this article we focus on the electromagnetic fields created by the quantum sources in vacuum. Also, since quantum diffusion does not  depend on spin, we replace the nucleons with scalar sources satisfying the Klein-Gordon equation. We defer analysis of the spin contribution and medium effects to future work. 

It is important to emphasize, that unlike the typical situation in  scattering theory, we cannot approximate the nucleon wave function by a plane wave, i.e.\ a state with a definite momentum. The 
collision theory usually considers  scattering of two wave packets that have a spatial extent much greater than the interaction range because the former is of macroscopic origin, while the latter is of a microscopic scale. This is one of the reasons that the incident and outgoing particles can be treated as plane waves \cite{Messiah}. Our situation is different because both scales are microscopic. The very fact that a collision at impact parameter $b$ happened implies that the spatial dimension of the wave packet after the collision is of order $b$, which is comparable with the interaction range. If we were interested in the short distance processes, we could have used the plane approximation because such a process probes only a small part of the wave function in momentum space. Since we are interested in the long-range electromagnetic fields, the plane wave approximation cannot be made.

The paper is organized as follows: In \sec{sec:b} we review the classical calculation of the electromagnetic field of a point particle that has been relied upon in the literature thus far.  In \sec{sec:d} we calculate the electromagnetic field strength using non-relativistic quantum mechanics and discuss two limits: the classical limit that yields the results of \sec{sec:b} and the plane wave limit of  quantum scattering theory in which the fields vanish altogether. Although the non-relativistic approximation is not valid for the relativistic collisions we are interested in, it provides a simple illustrative model of  quantum diffusion.  In \sec{sec:h} we calculate the fields in the ultra-relativistic approximation. The results show a strong quantum (vacuum) diffusion effect  that leads to the transverse expansion of the electromagnetic charge distribution. As a result, the electromagnetic fields are present  in the interaction region for times of the order of the system size. In  \sec{sec:k} we support our conclusions by numerically computing the electromagnetic fields. In \sec{sec:m} we present a brief summary and discussion of our results.

\section{Electromagnetic field of a classical source}\label{sec:b}

\begin{figure}[ht]
\begin{tabular}{cc}
      \includegraphics[height=6cm]{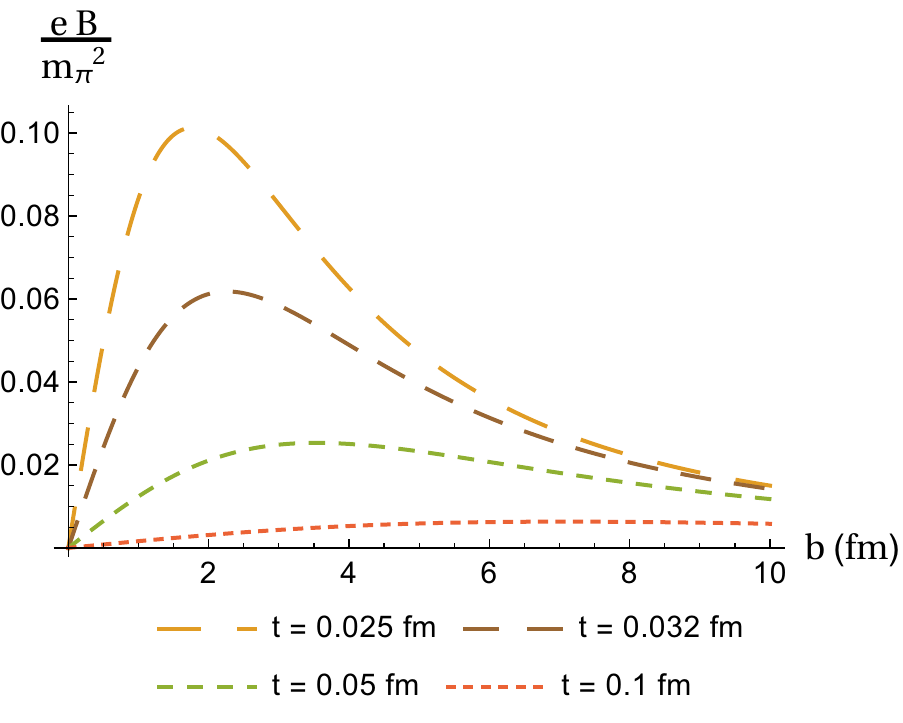} &
      \includegraphics[height=6cm]{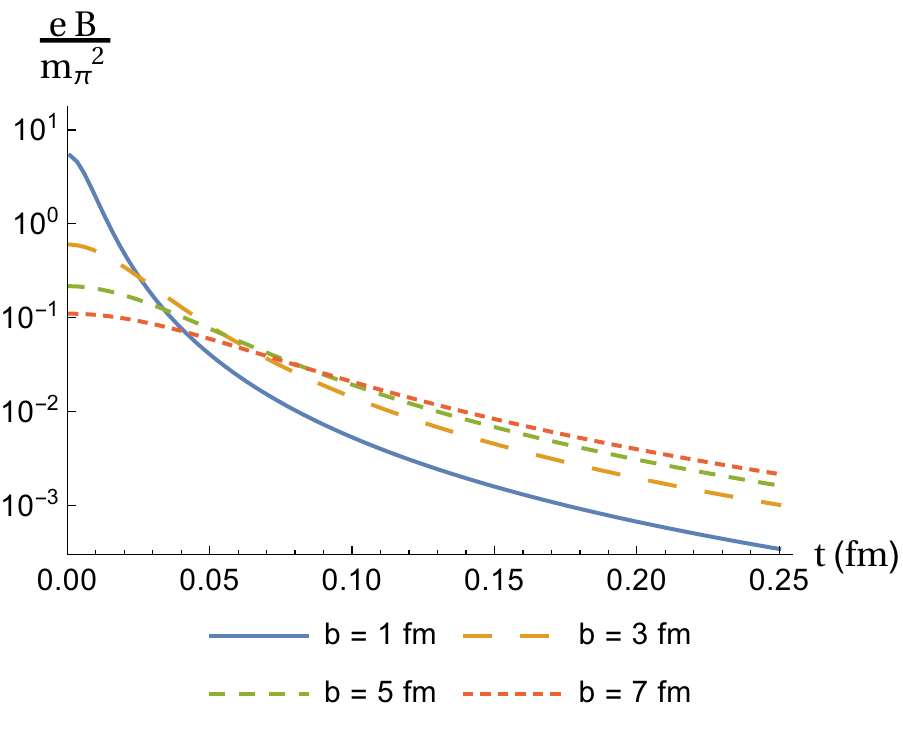}
      \end{tabular}
  \caption{Magnetic field $B$ created by a classical point charge as a function of impact parameter $b$ (left panel) and time $t$ (right panel),  see \eq{b15}.  }
\label{Classical-B}
\end{figure}

It is instructive to begin with the calculation of an electromagnetic field created by a classical point charge $e$. In the charge's rest frame, marked by the subscript ``0" throughout the paper, the charge and current densities are 
\ball{b1}
\rho_0(\b r_0,t_0)= e\delta(\b r_0)\,,\quad \b j_0(\b r_0,t_0)=0\,.
\gal
The corresponding potentials read
\ball{b3}
\varphi_0(\b r_0,t_0)= \frac{e}{4\pi r_0}\,,\quad \b A_0(\b r_0,t_0)=0\,.
\gal
Transitioning to the center-of-mass (COM) frame where the charge moves with the constant velocity $v$ along the $z$-axis  is accomplished by the transformation
\bal
&z_0= \gamma(z-vt)\,,\quad t_0=\gamma(t-v z)\,,\label{b5}\\
&A_{z0}= \gamma(A_z-v\varphi)\,,\quad \varphi_0=\gamma(\varphi-v A_z)\label{b6}\,.
\gal
The transverse components of the potentials do not change.  Denoting the transverse components of the position vector by $\b b$, so that  $\b r= z\unit z+\b b$, we obtain  
\ball{b8}
\varphi(\b r, t) = \frac{e}{4\pi}\frac{\gamma}{\sqrt{b^2+ \gamma^2(z-vt)^2}}\,,\quad \b A(\b r, t) = \frac{e}{4\pi}\frac{\gamma \b v}{\sqrt{b^2+ \gamma^2(z-vt)^2}}\,.
\gal
Alternatively, these equations can be obtained directly from the wave equations 
\ball{b10}
(\partial_t^2- \nabla^2)\b A= \b j\,,\quad (\partial^2_t - \nabla^2)\varphi= \rho\,,
\gal
with the potentials satisfying the Lorentz gauge condition
\ball{b12}
\b \nabla\cdot \b A+\partial_t \varphi=0\,,
\gal
and the current given by 
\ball{b14}
\rho(\b r, t) = e\delta(z-vt)\delta (\b b)\,,\quad \b j(\b r, t)= e\b v\delta(z-vt)\delta(\b b)\,.
\gal
The electromagnetic fields are calculated as
\bal
\b B&= \b \nabla \times \b A =  \frac{\gamma e v \unit \phi}{4\pi}\frac{b}{(b^2+\gamma^2(vt-z)^2)^{3/2}}\,,\label{b15}\\
\b E&= -\partial_t\b A-\b \nabla \varphi = \frac{\gamma e }{4\pi}\frac{\b b + (z-vt)\unit z}{(b^2+\gamma^2(vt-z)^2)^{3/2}}\,.\label{b16}
\gal 
where $\phi$ is the polar angle of the cylindrical coordinate system and $\unit\phi$ is the corresponding unit  vector.  Magnetic field \eq{b15} at $z=0$ is plotted in \fig{Classical-B}  for the future reference.

Eqs.~\eq{b15},\eq{b16} hold in the eikonal approximation. Namely, we assumed that the particle trajectory is a straight line. As explained in the Introduction, this is a valid approximation if the particle's deflection in the course of the collision is negligible.  The deflection is caused primarily by the strong nuclear force which does work $\sim\as/b$  on a particle moving along the $z$-axis. It can be neglected  compared to the particle's kinetic energy if $b\gg \as/(m\gamma)$, which holds for all relevant distances.

The classical approach of this section assumes that the particle trajectory is known: its position is fixed by the delta-functions in \eq{b14} and its momentum is $\b p= m\b v$. This is a good approximation far away from the charges. However, at small $b$ it breaks down giving way to the quantum treatment which we discuss in the next section.

\section{Non-relativistic Quantum source}\label{sec:d}

The electromagnetic current of quantum sources depends on their wave function. In this paper we model the wave function by a Gaussian of width $a$ in the particle's rest frame. The main advantage of this model is its simplicity. In particular, we can derive analytical results that clearly display the effects we are studying and serve as a benchmark for future calculations. Thus, suppose that at the initial (collision) time $t=0$, the wave function in the particle's rest frame  is given by 
\ball{d0}
\psi(\b r_0,0)= \frac{1}{\pi^{3/4} a^{3/2}}e^{-\frac{r_0^2}{2a^2}}\, ,
\gal
The corresponding wave function in  momentum space  reads
\ball{d1}
\psi_{\b k}(0)= \left(\frac{a^2}{\pi \hbar^2}\right)^{3/4}
e^{-\frac{a^2k^2}{2\hbar^2}}\,.
\gal



It is instructive to begin with a non-relativistic approximation, which applies if the valence charge is heavy. The time-dependence of the wave function \eq{d0} is given by 
\ball{d4}
\psi(\b r_0, t_0)&= \int \frac{d^3k}{(2\pi \hbar)^{3/2}} e^{\frac{i\b k\cdot \b r_0}{\hbar}}e^{-\frac{ik^2t_0}{2m\hbar }}\, \psi_{\b k}(0)=\frac{1}{\pi^{3/4}(a+i\hbar t_0/ma)^{3/2}}e^{-\frac{r_0^2}{2(a^2+i\hbar t_0/m)}}\,. 
\gal
The charge and current density in the rest frame can be computed as 
\bal
&\rho_0= e|\psi(\b r_0,t_0)|^2=\frac{e}{\pi^{3/2}(a^2+(\hbar t_0/ma)^2)^{3/2}}
e^{-\frac{r_0^2}{a^2+(\hbar t_0/ma)^2}}\,,\label{d6}\\
&\b j_0= \frac{e\hbar}{2mi}(\psi^*\b\nabla\psi-\psi\b\nabla\psi^* )=\frac{e\hbar^2 t_0 \b r_0}{m^2a^2\pi^{3/2}(a^2+(\hbar t_0/ma)^2)^{5/2}}e^{-\frac{r_0^2}{a^2+(\hbar t_0/ma)^2}}\,.
\gal
 Comparing this with the classical point source \eq{b1} we observe that (i) the charge density is spread over the region of size $a$ and  (ii) it is time-dependent even in the rest frame. 

The wave function in the COM frame is obtained by means of the Galilean transformation  
\bal
&z_0= z-vt\,,\quad t_0=t\,,\label{d8}\\
& j_{0z}= j_z-v \rho\,,\quad   \rho_0=  \rho\,. \label{d8a}
\gal
As a result, the charge and current densities of a particle moving with velocity $\b v= v\unit z$ read
\bal
&\rho(\b r, t)=\frac{e}{\pi^{3/2}(a^2+(\lambdabar t/a)^2)^{3/2}}
e^{-\frac{b^2+ (z-vt)^2}{a^2+(\lambdabar t/a)^2}}\,,\label{d9a}\\
&\b j(\b r, t)= \b v \rho(\b r, t)+ \frac{e\lambdabar^2 t [\b b+ (z-vt)\unit z]}{a^2\pi^{3/2}(a^2+(\lambdabar t/a)^2)^{5/2}}e^{-\frac{b^2+ (z-vt)^2}{a^2+(\lambdabar t/a)^2}}\, , \label{d9b}
\gal
where $\lambdabar= \hbar/m$ is the Compton wavelength. The classical limit of \eq{d9a},\eq{d9b} is reached by first setting $\lambdabar \to 0$, which yields 
\bal
& \rho(\b r, t)|_{\lambdabar \to 0}= \frac{1}{a^3\pi^{3/2}}e^{-\frac{(z-p_z t/m)^2+b^2}{a^2}}\,, \label{d11}\\
& {\b j}(\b r, t)|_{\lambdabar \to 0}= \frac{ev}{a^3\pi^{3/2}}e^{-\frac{(z-p_z t/m)^2+b^2}{a^2}}\,\unit z\,,\label{d12}
\gal
followed by the limit $a\to 0$. This procedure recovers the expression for the classical sources \eq{b14}.\footnote{Note that the limits $\hbar \to 0$ and $a\to 0$ do not commute. The classical result \eq{b14} is obtained by first taking $\hbar \to 0$ and then $a\to 0$ in which case both the momentum and the coordinate have definite values. This is why we cannot put $a$ to zero right away. } 

In  quantum scattering theory, the incoming and outgoing particles are described by states with definite momentum, which corresponds to the limit $a\to \infty$. This approximation is valid only when the uncertainty of particle position $a$ is much larger than the interaction range. This is clearly not the case in the problem we are discussing here.

The electromagnetic  retarded potentials in the rest frame  are
\bal
 &\varphi_0(\b r_0,t_0)= \frac{1}{4\pi}\int \frac{ \rho_0(\b r',t_0-|\b r_0-\b r'|)}{|\b r_0-\b r'|}d^3r'\,,  \label{g5}\\
 & \b A_0(\b r_0,t_0)=  \frac{1}{4\pi}\int \frac{ \b j_0(\b r',t_0-|\b r_0-\b r'|)}{|\b r_0-\b r'|}d^3r'  \,.  \label{g6}
\gal
  The corresponding electric field is 
\ball{g6E}
\b E_0(\b r_0, t_0)&= \int  \left\{ \frac{\rho_0(\b r',t') \b R}{R^3}+\frac{\b R}{R^2}\frac{\partial \rho_0(\b r',t') }{\partial t'}-\frac{1}{R}\frac{\partial {\b j_0}(\b r',t')}{\partial t'}\right\} d^3r'\,,
\gal
where $\b R= \b r_0-\b r'$ and $t'= t_0- R$.  Due to the spherical symmetry in the rest frame, there are only two non-vanishing potentials $\varphi_0(r_0,t_0)$ and $A_{r0}(r_0,t_0)$; magnetic field $\b B_0=0$ vanishes and the only non-vanishing component of electric field is $E_{r0}(r_0,t_0)$.

 At  early times, $t\ll a^2/\lambdabar$, the quantum diffusion can be neglected, which allows us to analytically determine the field strength. In this approximation  $\rho_0$ is time-independent and $\b j_0\approx 0$ so that  the integral in \eq{g5} can be done analytically. To this end we expand the Coulomb potential in a complete set of the  spherical harmonics using the well-known formula
\ball{g7}
\frac{1}{|\b r-\b r'|}= 4\pi \sum_{\ell=0}^\infty\sum_{m=-\ell}^\ell \frac{1}{2\ell+1}\frac{r_<^\ell}{r_>^{\ell+1}} Y_{\ell m}^*(\theta',\phi')Y_{\ell m}(\theta,\phi) \,,
\gal
and, after a simple integration, derive
\ball{g9}
\varphi_0(\b r_0,t_0)= \frac{1}{4\pi}\frac{e}{r_0}\text{erf}\left(\frac{r_0}{a}\right)\,.
\gal
Thus, in the COM frame
\ball{g11}
\varphi(\b r, t) = \frac{1}{4\pi}\frac{e\gamma}{\sqrt{b^2+(z-vt)^2}}\text{erf}\left(\frac{\sqrt{b^2+(z-vt)^2}}{a}\right)\,.
\gal
When $z-vt=0$ and $b\ll a$ we get 
\ball{g12}
\varphi_0\to \frac{e}{4\pi}\frac{2}{a\sqrt{\pi}}\,.
\gal
 As we expected, the parameter $a$ regulates the maximal strength of the field. The total electromagnetic energy of the charge is also finite and in the rest frame  at early times is given by  
\ball{g13}
U_0= \frac{1}{2}\int \rho_0 (\b r_0,t_0)\varphi_0(\b r_0,t_0)d^3r_0= \frac{1}{2(2\pi)^{3/2}}\frac{e^2}{a}\,.
\gal
In the limit of a classical point source $a\to 0$, \eq{g11} becomes a boosted Coulomb potential. In the opposite limit of definite momentum $a\to \infty$, when the particle is described by a plane wave, the electromagnetic field vanishes. 

At later times the quantum diffusion becomes very important. In particular, the electromagnetic current in the transverse, viz.\ $\unit b$, direction violates  relativistic causality. This compels us to do the full relativistic calculation which is described in the next section.

\section{Relativistic quantum sources}

In the relativistic case, the wave function in the rest frame in the coordinate representation  reads 
\bal
\psi(\b r_0, t_0)&= \int \frac{d^3k}{(2\pi \hbar)^{3/2}} e^{\frac{i\b k\cdot \b r_0}{\hbar}}e^{-\frac{i\e_k t_0}{\hbar }}\, \sqrt{\frac{m}{\e_k}}\psi_{\b k}(0)\label{h1}\\
& =  \frac{m^2}{(2\pi\hbar)^{3/2}}\left( \frac{a^2}{\pi \hbar^2}\right)^{3/4}\frac{4\pi \hbar}{r_0}\int_0^\infty \frac{d\kappa \kappa }{(1+\kappa^2)^{1/4}}e^{-\frac{i}{\lambdabar}t_0 \sqrt{1+\kappa^2}}e^{-\frac{a^2\kappa^2}{2\lambdabar^2}}\sin\frac{\kappa r_0}{\lambdabar} \,,\label{h2}
\gal
where $\e_k= \sqrt{m^2+k^2}$, $k= m\kappa$ and $\lambdabar= \hbar/m$. Eq.~\eq{h2} satisfies the Klein-Gordon equation and replaces the non-relativistic equation \eq{d4}. It is difficult to handle \eq{h2} analytically. To obtain a qualitative picture of the relativistic effects we first analyze the ultra-relativistic limit.

\subsection{Ultra-relativistic limit}\label{sec:h}

In the ultra-relativistic limit, $k\gg m$, one can derive a compact analytical result for the wave function. Eq.~\eq{h2} reduces to 
\ball{h4}
\psi(\b r_0, t_0)= \frac{m^2}{(2\pi\hbar)^{3/2}}\left( \frac{a^2}{\pi \hbar^2}\right)^{3/4}\frac{4\pi \hbar}{r_0}\int_0^\infty d\kappa \sqrt{\kappa}\, e^{-\frac{i}{\lambdabar}t_0 \kappa}e^{-\frac{a^2\kappa^2}{2\lambdabar^2}}\sin\frac{\kappa r_0}{\lambdabar}\,,
\gal
which can be integrated employing the formulas 3.462.1 and 9.240 from \cite{RG}.  We derive
\ball{h6}
\psi(\b r_0, t_0)&=\frac{i}{2^{3/4}\pi^{5/4} \sqrt{\lambdabar}\,r_0}
\left\{  \Gamma(3/4)\left[ 
- \Phi\left(\frac{3}{4};\frac{1}{2};-\frac{(r_0-t_0)^2}{2 a^2}\right)+  \Phi\left(\frac{3}{4};\frac{1}{2};-\frac{(r_0+t_0)^2}{2   a^2}\right)\right]
   \right.  \nonumber\\
  -& 
\left. 
\frac{i\sqrt{2}}{a}  \Gamma (5/4) \left[(r_0-t_0)
  \Phi \left(\frac{5}{4};\frac{3}{2};-\frac{(r_0-t_0)^2}{2 a^2}\right)+(r_0+t_0) 
  \Phi\left(\frac{5}{4};\frac{3}{2};-\frac{(r_0+t_0)^2}{2 a^2}\right)\right]
   \right\}\,,
\gal
where $\Phi$ is the confluent hypergeometric function. 
Derivatives of $\psi$ can be calculated using the same formulas by first differentiating the integrand of \eq{h4}. The relativistic expressions for the charge and current densities read
\ball{h10}
\rho = \frac{ie\hbar}{2m}[\psi^* \partial_t\psi - (\partial_t\psi^*)\psi]\,,\quad 
\b j= \frac{e\hbar}{2mi}[\psi^* \b\nabla\psi - \psi \b \nabla\psi^*]\,.
\gal
For example, in the rest frame the charge density is given by
\beql{h12}
\begin{split}
\rho(t_0, r_0) = \frac{e \Gamma \left( \frac{3}{4} \right) \Gamma \left( \frac{5}{4} \right)}{12 a^3 \pi^{5/2} r_{0}^{2}} 
\bigg\{ \left[ \Phi \left( \frac{3}{4}, \frac{1}{2}; \frac{-(t_0 - r_0)^2}{2 a^2} \right) - \Phi \left( \frac{3}{4}, \frac{1}{2}; \frac{-(t_0 + r_0)^2}{2 a^2} \right) \right] \times \\
\times \bigg( 6 a^2 \left[ \Phi \left( \frac{5}{4}, \frac{3}{2}; \frac{-(t_0 - r_0)^2}{2 a^2} \right) - \Phi \left( \frac{5}{4}, \frac{3}{2}; \frac{-(t_0 + r_0)^2}{2 a^2} \right)
 \right]+ \\
 + 5 \left[ (t_0 + r_0)^2 \Phi \left( \frac{9}{4}, \frac{5}{2}; \frac{-(t_0 + r_0)^2}{2 a^2} \right) + (t_0 - r_0)^2 
\Phi \left( \frac{9}{4}, \frac{5}{2}; \frac{-(t_0 - r_0)^2}{2 a^2} \right) \right] \bigg) + \\
+ 9 \left[ (t_0 - r_0) \Phi \left( \frac{5}{4}, \frac{3}{2}; \frac{-(t_0 - r_0)^2}{2 a^2} \right) -
 (t_0 + r_0) \Phi \left( \frac{5}{4}, \frac{3}{2}; \frac{-(t_0 + r_0)^2}{2 a^2} \right) \right] \times \\
\times \left[ (t_0 - r_0) \Phi \left( \frac{7}{4}, \frac{3}{2}; \frac{-(t_0 - r_0)^2}{2 a^2} \right) -
 (t_0 + r_0) \Phi \left( \frac{7}{4}, \frac{3}{2}; \frac{-(t_0 +r_0)^2}{2 a^2} \right) \right]\bigg\}.
\end{split}
\eeq
The current density can be computed in the same way. Eq.~\eq{h12} describes a traveling wave of the charge distribution. The source of this wave is the quantum diffusion. The diffusion wave travels with the speed of light, as is expected in the ultra-relativistic approximation. Recall, that \eq{h12} is the charge distribution of the particle at rest.  Transition to the COM frame  where the particle moves with velocity $v$ is accomplished using the Lorentz transformations \eq{b5},\eq{b6}.
The maximum of the charge distribution occurs when  
\bal
r_0-t_0= \sqrt{b^2+\gamma^2(z-vt)^2}-\gamma(t-vz)=0\,,
\gal
which implies $b= \sqrt{t^2-z^2}$. An implication of this result is that even in the midrapidity plane ($z=0$), the charge and current densities are large as late as  $t\sim R$, where $R$ is the system size.

Since $\Phi$ is  suppressed at large values of its argument, the main contribution to the potential arises from the region $|r_0\pm t_0|\ll \sqrt{2}a$ where $\Phi\approx 1$. Expanding \eq{h12} we obtain
\ball{h14}
\rho_0(\b r_0,t_0)\approx  \frac{\Gamma(5/4)\Gamma(7/4)}{2\pi^{5/2}}\frac{e}{a^3}\,.
\gal
The corresponding potential  is calculated as in \eq{g5}. At short distances, $r_0\to 0$, we get
\ball{h16}
\varphi_0(\b r_0,t_0)\approx \int_0^{\sqrt{2}a}\frac{e\Gamma(5/4)\Gamma(7/4)}{2\pi^{5/2}a^3}rdr
=\frac{\Gamma(5/4)\Gamma(7/4)}{2\pi^{5/2}}\frac{e}{ a}\approx \frac{0.3}{4\pi}\frac{e}{a}\,,\quad r_0\to 0\,.
\gal
In the COM frame 
\ball{h17}
\varphi(\b r,t)\approx\frac{0.3}{4\pi}\frac{e\gamma}{a}\,, \quad r\to 0\,.
\gal
As in the non-relativistic case, the potential is regulated at small distances by $a$, and is boosted by $\gamma$ as expected from its transformational properties under the boosts, see \eq{b6}. 

\subsection{Numerical solution}\label{sec:k}

\begin{figure}[ht]
\begin{tabular}{cc}
      \includegraphics[height=6cm]{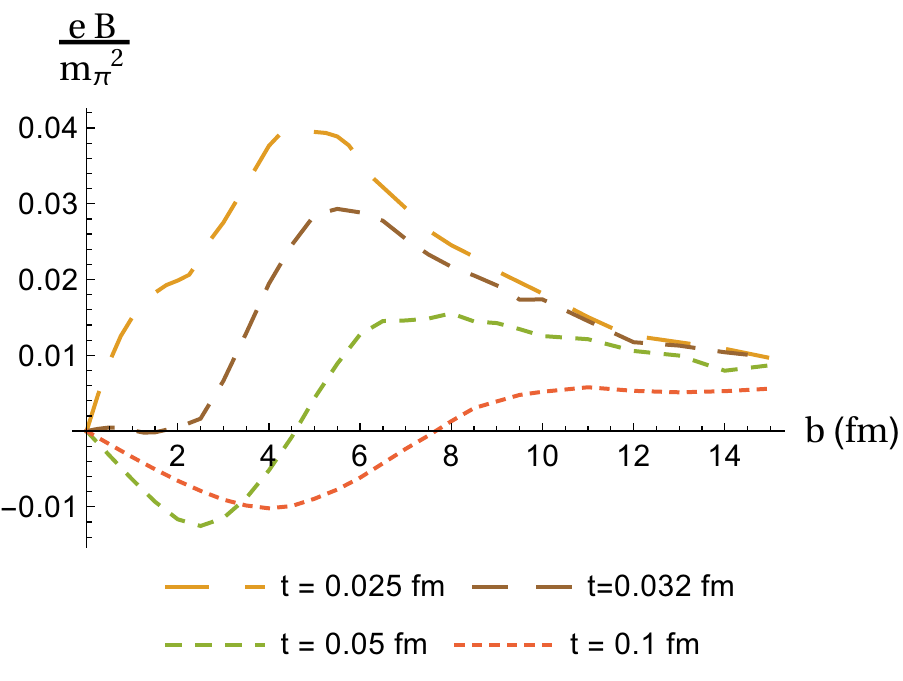} &
      \includegraphics[height=6cm]{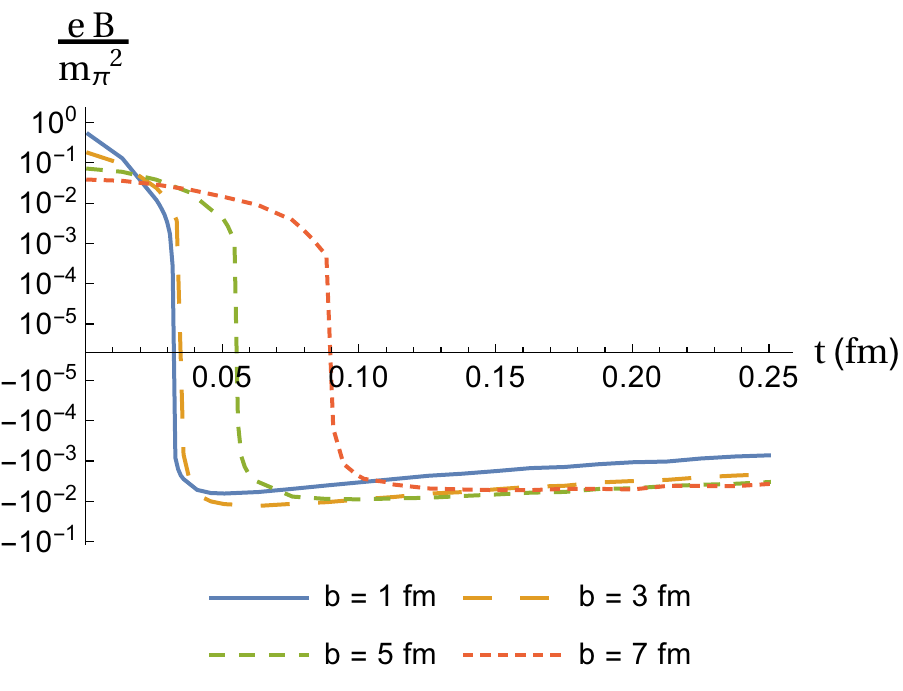}
      \end{tabular}
  \caption{Magnetic field  $B$ created by a quantum charge vs impact parameter $b$ (left panel) and time $t$ (right panel). }
\label{Quantum-B}
\end{figure}

Since the fields created by a relativistic quantum source are not amenable to an analytical solution, we calculated them numerically.  We present the results  in the cylindrical coordinate system span by the unit vectors $\unit b$, $\unit \theta$ and $\unit z$ in which  $\b B = B\unit \theta$, $\b E= E_\bot \unit b +   E_{\parallel}\unit z$.  The sources move with a constant velocity $v$ in the $+z$ direction, starting at the origin at $t = 0$.  Due to how the electromagnetic fields transform, in both the classical and relativistic case $E_{\perp} = B / v$, $E_\parallel = (z-vt)\b E_\bot \cdot \unit b/b$, see e.g.\ \eq{b15} and \eq{b16}, and so are not plotted here. For the purpose of the numerical calculation, the width of the Gaussian is fixed at $a$=1~fm, corresponding to the proton size,  $m=300$~MeV corresponding to the constituent quark mass and the collision energy  is $\gamma=100$. The results of our calculation are exhibited in \fig{Quantum-B}. In every figure we are looking at the midrapidity plane  $z = 0$.  

At very early times, the classical \fig{Classical-B} and quantum \fig{Quantum-B} calculations look qualitatively similar. The magnitude of the field of the classical sources is larger because the charge distribution of quantum sources is spread over the larger volume. As one can see in \fig{Quantum-B}, `very early times' is before about $t=0.05$~fm, which should be compared to the characteristic time 0.2~fm   when the nuclear color fields loose coherence. 

At later times, the classical and quantum calculations look very different. The magnitude of the field of the quantum sources at $t=0.25$~fm is an order of magnitude larger than the classical ones, which is indicative of the transversely traveling diffusion wave. 
Furthermore, unlike the fields of the classical sources, all components of fields of the quantum sources change sign at a certain $b$ ($t$). This indicates the position (time) at which most of the charge and current density effectively passes the observation point. To wit, initially most of the charge and current density is localized within a sphere of radius $a$, so that an observer in the charge rest frame located at $b>a$ (and $z=0$) sees  all charge near the origin. At later times most of the charge diffuses outside a sphere of radius $b$, so that the observer will see most of the charge located on the other side with respect to the origin. The negative sign of magnetic field at small $b$'s and large $t$'s is due to the second and third terms in \eq{g6E}, which imply that the diffusion current increases while the charge density decreases with time in that region.  The fact that the fields flip their direction may have  important implications for the interpretation of the charge separation effect \cite{Kharzeev:2007tn}.

\section{Summary and outlook}\label{sec:m}

The classical scattering theory deals with  point particles that have definite position and momentum, while the quantum scattering theory usually deals with asymptotic states with definite momentum. So far calculations of the electromagnetic field produced in relativistic  collisions have been done in the classical approach which plainly contradicts the uncertainty relation.  On the other hand, treating particles as plane waves is also not a reasonable approach because we are interested in the fields created in a microscopic volume. Thus, we have to deal with wave packets of finite size that   were described in this paper by a Gaussian of width $a$ (in the coordinate space). In the limit $a\to 0$ we recovered expressions for the Coulomb field of a classical point charge, while at $a\to \infty $ (the plane wave limit) the electromagnetic field vanishes. 

The time-evolution of the wave function of the source depends on the dispersion relations. While some analytical results can be derived in the non-relativistic and ultra-relativistic approximations, the most accurate approach is to use the exact relativistic dispersion relation which has been done numerically. Our main conclusion is that  that quantum treatment is absolutely essential in obtaining realistic electromagnetic fields. This is true not only at short distances $\sim 1$~fm but even at larger distances in the entire interaction volume. The electromagnetic fields produced by the quantum sources stay longer in the interaction volume than their classical counterparts and have a significantly different spatial distribution. 

Since our primary interest  in this work was studying the time evolution of the quantum sources, we neglected the spin and medium effects. However, a phenomenologically sound approach must necessarily take  these effects into account. These will be addressed elsewhere.

\acknowledgments
We  are grateful to James Vary, Xingbo Zhao, Guangyao Chen and Yang Li  for informative discussions during our weekly group meetings.  This work  was supported in part by the U.S. Department of Energy under Grant No.\ DE-FG02-87ER40371.



\end{document}